\documentclass[twocolumn,showpacs,aps]{revtex4}
\usepackage{amsmath,amsthm,amscd}
\usepackage{dcolumn}
\usepackage{graphicx}
\usepackage{epsfig}
\begin{document}
\newcommand{\xh}{\ensuremath{\hat{\mathbf{x}}}}
\newcommand{\yh}{\ensuremath{\hat{\mathbf{y}}}}
\newcommand{\zh}{\ensuremath{\hat{\mathbf{z}}}}
\newcommand{\xha}{\ensuremath{\hat{\mathbf{x}}_1}}
\newcommand{\xhb}{\ensuremath{\hat{\mathbf{x}}_2}}
\newcommand{\xhc}{\ensuremath{\hat{\mathbf{x}}_3}}
\newcommand{\rb}{\ensuremath{\mathbf{r}}}
\newcommand{\q}{\ensuremath{\mathbf{Q}}}
\newcommand{\Rb}{\ensuremath{\mathbf{R}}}
\newcommand{\wh}{\ensuremath{\hat{w}}}
\newcommand{\xhi}{\ensuremath{\hat{\mathbf{x}}_i}}
\newcommand{\xhj}{\ensuremath{\hat{\mathbf{x}}_j}}
\newcommand{\xthi}{\ensuremath{\tilde{\hat{\mathbf{x}}}_i}}
\newcommand{\xthj}{\ensuremath{\tilde{\hat{\mathbf{x}}}_j}}
\newcommand{\kb}{\ensuremath{\mathbf{k}}}
\newcommand{\Qb}{\ensuremath{\mathbf{Q}}}
\newcommand{\Sb}{\ensuremath{\mathbf{S}}}
\newcommand{\Hb}{\ensuremath{\mathbf{H}}}
\newcommand{\Db}{\ensuremath{\mathbf{D}}}
\newcommand{\db}{\ensuremath{\mathbf{d}}}
\newcommand{\pb}{\ensuremath{\mathbf{p}}}
\newcommand{\Pb}{\ensuremath{\mathbf{P}}}
\newcommand{\Vb}{\ensuremath{\mathbf{V}}}
\newcommand{\Wb}{\ensuremath{\mathbf{W}}}
\newcommand{\deltab}{\ensuremath{\mathbf{\Delta}}}
\newcommand{\Ss}{\ensuremath{\mathcal{S}}}
\newcommand{\As}{\ensuremath{\mathcal{A}}}
\newcommand{\Cs}{\ensuremath{\mathcal{C}}}
\title{\textbf{Canted-spin-caused electric dipoles: a local symmetry theory}}
\author{T. A. Kaplan and S. D. Mahanti}
\affiliation{Department of Physics \& Astronomy, Michigan State University\\
East Lansing, MI 48824}

\begin{abstract}
A pair of magnetic atoms with canted spins $\mathbf{S}_a,\mathbf{S_b}$ can give rise to an electric dipole
moment $\Pb$. Several forms for the behavior of such a moment have appeared in the theoretical literature, some
of which have been invoked to explain experimental results found in various multiferroic materials. The forms
that require canting of the spins are $\Pb_1\propto\Rb\times(\Sb_a\times\Sb_b), \Pb_2\propto\Sb_a\times\Sb_b$,
and $\Pb_3\propto\Sb_a\Rb\cdot\Sb_a-\Sb_b\Rb\cdot\Sb_b$, where $\Rb$ is the relative position of the atoms and
$\Sb_a,\Sb_b$ are unit vectors. To unify and generalize these various forms we consider $\Pb$ as the most
general quadratic function of the spin components that vanishes whenever $\Sb_a$ and $\Sb_b$ are collinear, i.e.
we consider the most general expressions that require spin canting. The study reveals new forms. We generalize
to the vector $\Pb$, Moriya's symmetry considerations regarding the (scalar) Dzyaloshinskii-Moriya energy
$\Db\cdot \Sb_a\times\Sb_b$ (which led to restrictions on $\Db$). This provides a rigorous symmetry argument
which shows that $\Pb_1$ is allowed no matter how high the symmetry of the atoms plus environment, and gives
restrictions for all other contributions. The analysis leads to the suggestion of terms omitted in the existing
microscopic models, suggests a new mechanism behind the ferroelectricity found in the `proper screw structure'
of CuXO$_2$, X=Fe,Cr, and predicts an unusual antiferroelectric ordering in the antiferromagnetically and
ferroelectrically ordered phase of RbFe(MoO$_4$)$_2$.
\end{abstract}
 \maketitle

\textbf{I. Introduction}

Great recent interest in multiferroic materials,
e.g.~\cite{kimura,hur,lawes,katsura,kenzelmann3,kimura2,yamasaki,mostovoy,sergienko,harris0,kaplan,arima0,jia,
cheong,radaelli,kenzelmann2,harris,ederer,seki0,kenzelmann,seki,nakajima,mostovoy2,harris2,choi,
malashevich,khomskii,soda,tokura,ishiwata, mochizuki2,choi2,chapon,sergienko2}, where magnetic ordering of
various sorts induces ferro- or ferri-electricity, forces one to understand the microscopic foundation for this
surprising, and possibly useful effect. Broadly, there are two sources of this fascinating effect. One, found in
many materials, depends on the canting of the spins in an essential way (often referred to as ``antisymmetric
dependence of the dipole moment on the spins").
\cite{kimura,hur,lawes,katsura,kenzelmann3,kimura2,yamasaki,mostovoy,sergienko,harris0,kaplan,arima0,jia,
cheong,radaelli,kenzelmann2,harris,ederer,seki0,kenzelmann,seki,nakajima,mostovoy2,harris2,choi,malashevich,
tokura, khomskii,soda,ishiwata, mochizuki2} The
other~\cite{jia,khomskii,ishiwata,mochizuki2,choi2,chapon,sergienko2} derives from ordering which may or may not
involve canted spins, i.e. any canting is incidental (``symmetric dependence"). For clarity of presentation, the
present paper deals exclusively with the case where canting is essential. This case embodies the meaning of our
term ``canted-spin-caused electric dipoles".

 One microscopic approach to this effect, due to
Katsura, Nagaosa and Balatzky (KNB)~\cite{katsura} is derived by considering a model containing a pair of
magnetic ions whose average spins $\Sb_a,\Sb_b$ are constrained to be in arbitrary directions. Such a constraint
is imagined to result from exchange and anisotropy fields originating from the long range ordered magnetic state
of the crystal. E.g, the magnetic state might be a spiral and the ion pair considered would be any neighboring
pair participating in the spiral (with canted spins). In~\cite{katsura} it is found that the electron density
becomes distorted by a combination of spin-orbit coupling $V_{SO}$ and interionic electron hopping $t$. To
leading order in $t$ and $V_{SO}$  an electric dipole moment is found, given by
\begin{equation}
 c\Rb\times(\Sb_a\times\Sb_b),\label{one}
\end{equation}
where $\Rb$ is the displacement of one ion relative to the other, and $c$ is a coefficient, discussed below.

Sergienko and Dagotto~\cite{sergienko} also considered a pair of magnetic atoms with canted spins, and noted
that the Dzyaloshinskii-Moriya (DM) term, $\Db\cdot\Sb_a\times\Sb_b$, in the superexchange energy also gave the
same form when the intervening oxygen ion was allowed to move off center. This is spoken of as spin-lattice
interaction, or magnetostriction.

A different approach is based on the complete crystal with spiral-like spin ordering; it has led to results
consistent with~(\ref{one}). A derivation in this vein based on spin-lattice interactions by Harris et
al~\cite{harris0} has been given; they consider magnetostriction both of the type coming from the DM coupling,
which originates in the antisymmetric part of the exchange tensor, and that coming from the symmetric part; see
also~\cite{mochizuki2}. There are also phenomenological derivations of magneto-ferroelectricity using symmetry
arguments via Landau theory~\cite{lawes,kenzelmann2,sergienko2,harris2}, and Landau-Ginzberg
theory~\cite{mostovoy}.

Also relevant here is a model~\cite{kaplan} that is closely related to the KNB approach, again involving a pair
of atoms, small hopping and spin-orbit coupling. In~\cite{kaplan}a the expression~(\ref{one}) was found, where
the assumption was made that the spatial symmetry of the situation was the symmetry of a pair of points in
space, an assumption also made in~\cite{katsura, sergienko}. However, in~\cite{kaplan}b, a lower symmetry was
studied, which led to the possibility of another component of the dipole, namely in the direction
\begin{equation}
 \Sb_a\times \Sb_b,\label{onep}
\end{equation}
thus questioning the generality of~(\ref{one}).\footnote{The lower symmetry caused by orbital ordering was
considered in~\cite{jia}, yet no additional terms like~(\ref{onep}) were found. An explanation of this apparent
dilemma can be found in Section II, Case 1, example (c).} This question was also raised, considering extended
systems, in~\cite{arima0} and~\cite{kenzelmann}. In~\cite{arima0}, experimental evidence in
CuFeO$_2$~\cite{nakajima} for this new possibility, occurring in the proper screw structure, was noted; a
symmetry argument based on the observed spiral was given~(\cite{arima0}), as well as a suggested microscopic
mechanism behind the observation (to be discussed further below). A similar situation was found in
CuCrO$_2$.~\cite{soda} In connection with~\cite{kenzelmann}, the question was answered in~\cite{kenzelmann2}
where it was shown by an experimental example, RbFe(MoO$_4)_2$ (RFMO), and a Landau theory analysis, that this
$\Sb_a\times \Sb_b$-component can exist.

In overlapping time frames, a paper by Jia et al~\cite{jia} followed the basic approach of Katsura et al,
considering a system with two magnetic atoms. In addition to giving a serious estimate of the coefficient $c$
in~(\ref{one}), more general considerations added to~(\ref{one}) two additional terms, one is the well-known
exchange striction (which doesn't concern us here because it doesn't require spin-canting) plus a new type of
term, proportional to
\begin{equation}
(\Rb\cdot\Sb_a)\Sb_a-(\Rb\cdot\Sb_b)\Sb_b,~\label{j}
\end{equation}
where $\Sb_a,\Sb_b$ are unit vectors. It is seen that this gives non-zero $\pb$ only if the spins are not
collinear, which conforms to our general idea, in fact the precise definition, of a `canted-spin-caused'
electric dipole. One notices that unlike the previous forms, which are bilinear in the 2 spins, this falls under
the general heading of being quadratic in the spins. Arima~\cite{arima0} refers to this result, and generalizes
it in a way that leads to a polarization parallel to the spiral wave vector $\textbf{Q}$ in a ``proper screw
structure" (a spiral where the spin plane is normal to $\Qb$). (Since $\Qb||\Rb$ in his case,~(\ref{j}) clearly
would give zero for such a spiral.) We will point out (in Section IV) a different microscopic mechanism that
also gives $\pb$ in the direction of $\Sb_a\times\Sb_b$, that may be responsible for the behavior observed in
the proper screw structure, and that also applies to RFMO (which is not a proper screw structure). (This
mechanism is linear in the spin-orbit coupling strength while Arima's is quadratic.)

Thus we see a veritable zoo of forms for the canted-spin caused dipole moment. One must ask, what others might
exist? A common theme in all those mentioned is that they are quadratic in the pair of spins. The theory
presented here considers the most general quadratic function that represents canted-spin-caused dipoles, and
analyzes various forms allowed under whatever symmetry is ``seen" by the pair of magnetic ions.~\cite{kaplanarxiv}  Since it
includes the cases already known, it represents a general unified picture of the possible forms. The theory is
model-independent and local (treating a single pair of magnetic ions or atoms). It is closely analogous to an
argument leading to the conditions on the DM vector $\mathbf{D}$ (Moriya's rules) imposed by the symmetries of
the magnetically disordered crystal.~\cite{moriya}

The results show that forms far more general than~(\ref{one}),(\ref{onep}), and (\ref{j}) are to be expected in
general, and which symmetries, or, rather, their absence, are required for the more general forms. The theory
also offers an explanation for the fact that~(\ref{one}) is found in many materials whereas the other forms have
been found in relatively few (as far as we are aware). The analysis leads to the suggestion of new terms omitted
from the microscopic theories. And it predicts an unusual antiferroelectric ordering in the
antiferromagnetically and ferroelectrically ordered phase of RbFe(MoO$_4$)$_2$.

To apply this local theory to solids, one must determine how $\pb$ for a single bond propagates through the
crystal. This is discussed through a few examples.

Section II reviews an analysis of the scalar quantity $\Db\cdot(\Sb_a\times\Sb_b)$ that derives symmetry
restrictions on the DM vector $\Db$ (Moriya's rules), and applies an analogous analysis to the dipole moment
$\pb$, which is of course a vector. Essential to the latter is expressing $\pb$ as a general homogenous
quadratic function of $\Sb_a$ and $\Sb_b$. This restriction is made in the spirit of leading order perturbation
theory treating the hopping, spin-orbit coupling, and/or magnetostrictive atomic displacements as small. It
applies to the approaches of KNB and related, as well as to the spin-lattice interaction approach
of~\cite{sergienko} and the corresponding work of Harris et al~\cite{harris0}, and to the problem of CuXO$_2$,
X=Fe,Cr~\cite{arima0,nakajima,soda}. Section III presents examples in crystals, some ideal, and some
corresponding to the structures of real multiferroic crystals. Section IV contains some concluding remarks.
Appendix I discusses the general bilinear function of 2 spins, with matrix $B$ of the  quadratic form for each
component of $\pb$. It shows that the most general spin-canted-caused dipole form originates from the
antisymmetric part of $B$, and is linear in $\Sb_a\times\Sb_b$. We also consider, in the text, the most general
\emph{quadratic} function of the spins, and find additional contributions to $\pb$, a special case of which is
of form~(\ref{j}). Thus the overall results generalize all known forms. Appendix 2 describes the simple
microscopic model~\cite{kaplan} and its application as a check on the results of the abstract model-independent
symmetry arguments.
\begin{figure}[h]
\centering\includegraphics[height=2in]{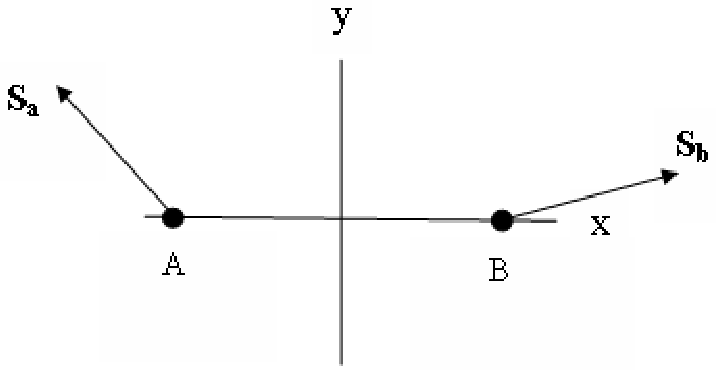}\vspace{-40pt}
 \caption{The coordinate system and an example of the two spins. The z-direction is out of the paper.}
\label{fig1}
\end{figure}\\
\textbf{II. Symmetry analysis of the electric dipole produced by two canted spins.}

 We begin by reviewing an argument leading to Moriya's rules.\footnote{Moriya~\cite{moriya} states ``the
rules are obtained easily"; he also gives an explicit formula for \textbf{D}. It is not clear if he obtained the
rules through his formula or some other way.} One considers the possible existence of a term in the energy of
the form $E_{DM}=\Db\cdot(\Sb_a\times\Sb_b)$, where $\Sb_a$ and $\Sb_b$ are the spins at sites A and B
respectively. $\Db$ is ``a constant vector", to quote Moriya~\cite{moriya}. Its sign obviously depends on the
(arbitrary) order chosen to write the spins in the cross-product. If one adheres to a choice, e.g. spin at
position A $\times$ spin at position B, then $\Db$ is a constant. I.e. it is a property of the structure,
atom-pair plus surroundings exclusive of magnetic ordering and spin-orbit coupling. One explores the conditions
imposed on $\Db$ by possible symmetries of the structure (without spin ordering), i.e. rotations which return
the two sites plus surroundings to itself, with the requirement that $E_{DM}$ be unchanged (as a term in a
Hamiltonian, it's a scalar under such operations). Important is the fact that $\Db$ is fixed in the structure
(as seen in Moriya's mathematical expression for it), so that $\Db$ is the same before and after the operation,
emphasizing again that the order of the spins remains, spin at A $\times$ spin at B.

As a first illustration, inversion about the coordinate origin O in Fig. 1 simply interchanges $\Sb_a$ and
$\Sb_b$, so that the new spin at site A, $\Sb_a^\prime=\Sb_b$, and $\Sb_b^\prime=\Sb_a$. Assuming inversion is a
symmetry of the structure, one concludes
$\Db\cdot\Sb_a\times\Sb_b=\Db\cdot\Sb_a^\prime\times\Sb_b^\prime=-\Db\cdot\Sb_a \times\Sb_b$ for arbitrary
$\Sb_a,\Sb_b$. Moriya's Rule 1 follows: Given this inversion symmetry, $\Db=0$. Next consider Rule 2. Suppose a
mirror plane perpendicular to AB passes through O. Then the transformed spins are
\begin{eqnarray}
\Sb_a^\prime&=&\xh S_{bx}-\yh S_{by}-\zh S_{bz}\nonumber\\
\Sb_b^\prime&=&\xh S_{ax}-\yh S_{ay}-\zh S_{az}\label{mirrorperp}
\end{eqnarray}
which yields
\begin{equation}
\Sb_a^\prime\times\Sb_b^\prime=-\hat{x}(\Sb_a\times\Sb_b)_x+\hat{y}(\Sb_a\times\Sb_b)_y
+\hat{z}(\Sb_a\times\Sb_b)_z~\label{2}
\end{equation}
Again, equating $\Db\cdot\Sb_a\times\Sb_b=\Db\cdot\Sb_a^\prime\times\Sb_b^\prime$ gives $D_x=0$ (Rule 2). This
procedure can be seen to yield all 5 rules.\footnote{We have used the axial-vector property of the spins; the
results are unchanged if they are considered vectors.}

Now consider the electric dipole moment $\pb$, a vector.\footnote{We find it convenient to use a notation
different from that in the abstract.} As motivated above, we consider $\pb$ caused by a pair of spins as the
general quadratic function
\begin{equation}
\pb=\sum_{\gamma,i,j,\nu,\mu}\hat{\gamma}B_{\gamma ij\nu\mu}S_{i\nu}S_{j\mu}\label{quad}
\end{equation}
where $\gamma, \nu$ and $\mu$ run over the Cartesian coordinates  $x,y,z, \hat{\nu}$ being the corresponding
unit vectors with $\hat{\nu}\cdot\hat{\mu}=\delta_{\nu\mu}$, and $i,j$ run over the site or spin labels, $a,b$.
We
consider separately the two cases, $i\ne j$ and $i=j$.\vspace{.1in} \\
\textbf{Case 1}: $\mathbf{i\ne j}$\\
(\ref{quad}) becomes
\begin{equation}
\pb=\sum_{\gamma,\nu,\mu}\hat{\gamma}B_{\gamma\nu\mu}S_{a\nu}S_{b\mu}
\end{equation}
where $B_{\gamma\nu\mu}=B_{\gamma a b\nu\mu}+B_{\gamma b a\mu\nu}.$
 In Appendix 1 it is shown that for this function (which is bilinear in the spins) to represent a canted-spin-caused
 dipole, it must
be a function of $\Sb_a\times\Sb_b\equiv\Wb$ that is linear homogeneous in $\Wb$, its most general form being
\begin{equation}
\pb=\sum \Cs_{\nu\mu}\hat{\nu}\hat{\mu}\cdot\Wb\equiv\stackrel{\Rightarrow}{\Cs}\cdot\Wb.\label{3}
\end{equation}
Here $\Cs_{\nu x}=B_{\nu yz}^a,\Cs_{\nu y}=B_{\nu zx}^a, \Cs_{\nu z}=B_{\nu xy}^a$, with
$B_{\gamma\nu\mu}^a=(B_{\gamma \nu\mu}-B_{\gamma \mu\nu})/2$.
 The form~(\ref{3}) also applies to the spin-lattice
mechanism via the DM term ($\stackrel{\Rightarrow}{\Cs}$ is related to the derivatives of the DM vector $\Db$ with respect to lattice distortions from the non-magnetic crystal structure).

The symmetric contribution, from $B_{\gamma\nu\mu}^s=(B_{\gamma\nu\mu}+B_{\gamma\mu\nu})/2$, is also important to multiferroics in general. But for simplicity, we focus in this paper on the canted-spin-caused part.

To connect with existing literature, we write $\Cs_{\nu\mu}=\Ss_{\nu\mu}+\As_{\nu\mu}$ where $\Ss$ and $\As$ are the symmetric and
antisymmetric parts of the matrix $\Cs$, allowing the separation of $\pb$ into the corresponding terms:
$\pb=\pb_\Ss+\pb_\As$.\footnote{There is a very different sense in which $\pb$ is written as a sum
$\pb_s+\pb_a$. Namely, in~\cite{mochizuki2} and elsewhere, $\pb_s$ is attributed to that obtained from the
spin-lattice interaction associated with the symmetric part of the exchange tensor, $\pb_a$ to the antisymmetric
part. In the present work, for the model of spin-lattice interaction, both $\pb_\Ss$ and $\pb_\As$ originate
from the antisymmetric part of the exchange tensor.} In particular, from~(\ref{3}) follows
\begin{eqnarray}
\pb_\As&=&\hat{x}(\As_{xy}W_y+\As_{xz}W_z)+\hat{y}(\As_{yx}W_x+\As_{yz}W_z)\nonumber\\
& &+\hat{z}(\As_{zx}W_x+\As_{zy}W_y).\label{4}
\end{eqnarray}
It is easily verified that this is
\begin{eqnarray}
\pb_\As&=&\db\times\Wb,\nonumber\\
\db&=& -(\hat{x}\As_{yz}+\hat{y}\As_{zx}+\hat{z}\As_{xy}).\label{5}
\end{eqnarray}
Thus we have connected to the important term~(\ref{one}), which is a special case of~(\ref{5}) in which
$\db=\db_{||}$, along $\Rb$, or $\hat{x}$ in the coordinate system of FIG. 1.

Recall that standard transformation theory in which we apply a rotation $\stackrel{\Rightarrow}{U}\equiv\sum
U_{\nu\mu}\hat{\nu}\hat{\mu}$ to~(\ref{3}) gives
\begin{equation}
\pb^\prime\equiv\stackrel{\Rightarrow}{U}\cdot\pb=\stackrel{\Rightarrow}{\Cs}^\prime\cdot\Wb^\prime,\nonumber
\end{equation}
where $\stackrel{\Rightarrow}{\Cs}^\prime=\sum \Cs_{\nu\mu}^\prime\hat{\nu}\hat{\mu}$, $\Cs^\prime=U\Cs U^{-1}$
and $\Wb^\prime=\stackrel{\Rightarrow}{U}\cdot\Wb(=\Sb_a^\prime\times\Sb_b^\prime)$. We consider $U$ as real and
unitary. When $U$ is a symmetry operation, as described above, the matrix $\Cs$ is unchanged, i.e.
$\Cs^\prime=\Cs$. Thus our fundamental equation for applying symmetry operations is
\begin{equation}
\pb^\prime=\sum\Cs_{\nu\mu}\hat{\nu}\hat{\mu}\cdot\Wb^\prime.\label{7}
\end{equation}
The  relation $\Cs^\prime=\Cs$ is analogous to $\Db$ being unchanged under a symmetry operation.~\cite{moriya}

We now apply rotations that leave the structure, sites A and B plus magnetically disordered environment,
unchanged, and require $\pb$ to satisfy its vector property. This requirement is applied for each of Moriya's
list of (five) rotations (all possibilities that take the sites A and B into themselves).\\
\textbf{1. Inversion through O} \\
As before $\Wb^\prime=-\Wb$. Thus~(\ref{7}) gives $\pb^\prime=-\pb$. This is precisely what a vector should do
under inversion. Thus \emph{inversion invariance gives no restriction on $\pb$}.\\
\textbf{2. Mirror $\perp$ AB}\\
The reflected $\Wb$ is given in~(\ref{2}): $(W_x^\prime,W_y^\prime,W_z^\prime)=(-W_x,W_y,W_z)$. Thus~(\ref{7})
becomes
\begin{eqnarray}
\pb^\prime=(-\Cs_{xx}W_x+\Cs_{xy}W_y+\Cs_{xz}W_z,\nonumber\\
-\Cs_{yx}W_x+\Cs_{yy}W_y+\Cs_{yz}W_z,\nonumber\\
-\Cs_{zx}W_x+\Cs_{zy}W_y+\Cs_{zz}W_z)\nonumber
 \end{eqnarray}
 The vector property says $\pb^\prime=(-p_x,p_y,p_z)$, with $p_\nu$ from~(\ref{3}). Therefore $\Cs$ must have the form
 \begin{equation}
 \Cs=\left(\begin{array}{ccc}
 C_{xx} & 0 & 0\\
 0 & C_{yy} & C_{yz}\\
 0 & C_{zy} & C_{zz}
 \end{array} \right).\label{8}
 \end{equation}
 We see that this symmetry requires the only contribution to $\As_{\nu\mu}$ be $\As_{yz}$. Thus ~(\ref{5}) gives
 $\db=-\hat{x}\As_{yz}$, i.e. $\db_{||}$, parallel (or antiparallel)
 to $\Rb$.\\
\textbf{3. Mirror includes AB}\\
We can take the mirror as the $xy$-plane. Since this involves no interchange of $\Sb_a$ and $\Sb_b$, $\Wb$
behaves as a pseudovector so $\Wb^\prime=(-W_x,-W_y,W_z)$. Then~(\ref{7}) reads
\begin{eqnarray}
\pb^\prime=(-\Cs_{xx}W_x-\Cs_{xy}W_y+\Cs_{xz}W_z,\nonumber\\
-\Cs_{yx}W_x-\Cs_{yy}W_y+\Cs_{yz}W_z,\nonumber\\
-\Cs_{zx}W_x-\Cs_{zy}W_y+\Cs_{zz}W_z).\nonumber
\end{eqnarray}
Comparing with the vector property $\pb^\prime=(p_x,p_y,-p_z)$ leads to the restricted form
\begin{equation}
\Cs=\left(\begin{array}{ccc}
0 & 0 & C_{xz}\\
0 & 0 & C_{yz}\\
C_{zx} & C_{zy} &  0 \end{array} \right).\label{9}
\end{equation}
This result implies $\db$ lies in the mirror plane.\\
\textbf{4. 2-fold rotation axis $\perp$ AB} \\
We can take this as the z-axis, so that $\Sb_a^\prime=(-S_{bx},-S_{by},S_{bz})$ and $a\leftrightarrow b$. This
gives
\begin{equation}
\Wb^\prime=(W_x,W_y,-W_z)\nonumber
\end{equation}
 Thus~(\ref{7}) becomes
\begin{eqnarray}
\pb^\prime=(\Cs_{xx}W_x+\Cs_{xy}W_y-\Cs_{xz}W_z,\nonumber\\
\Cs_{yx}W_x+\Cs_{yy}W_y-\Cs_{yz}W_z,\nonumber\\
\Cs_{zx}W_x+\Cs_{zy}W_y-\Cs_{zz}W_z).\nonumber
\end{eqnarray}
 Comparing with the vector property $\pb^\prime=(-p_x,-p_y,p_z)$ yields the same $\Cs$ as~(\ref{9}).
So this symmetry implies $\db\perp$ rotation axis.\\
\textbf{5. n-fold axis along AB, $n\ge2$}\\
Here $\Wb^\prime=(W_x,cW_y-sW_z,sW_y+cW_z)$, where $(c,s)\equiv (\cos\theta,\sin\theta),\theta=$ the rotation
angle. The vector property of $\pb$ demands $\pb^\prime=(p_x,c p_y-s p_z,s p_y+c p_z)$. We again equate this
expressed in terms of $\Wb$ (using~(\ref{3})) with the corresponding equation for $\pb^\prime$ given
by~(\ref{7}). For $n>2$ this leads to
\begin{equation}
\Cs=\left(\begin{array}{ccc}
C_{xx} & 0 & 0\\
0 & C_{yy} & C_{yz}\\
0 & -C_{yz} &  C_{yy} \end{array} \right)\ \mbox{for}\ n>2.\label{10}
\end{equation}
While this result is valid for all $n>2$, it changes for $n=2$, as
follows: The conditions $C_{zz}=C_{yy}$ and $C_{zy}=-C_{yz}$ no longer hold. The reason for the difference
between $n=2$ and $n\ne2$ is that for $n=2 (\theta=\pi)$, there is no mixing between $y-$ and $z-$ components,
unlike the case $n\ne2$. In either case, the form of $\Cs$ implies $\db=\db_{||}$. In contrast to the dipole moment $\pb$,
 it is interesting to note that the consequences of these symmetry
operations on the DM vector $\Db$ are independent of $n$.~\cite{moriya}

These results were checked against the microscopic model calculation in~\cite{kaplan} (see Appendix 2).

An important conclusion to be drawn from these results is that  the contribution to $\pb$ coming from
$\db_{||}\times(\Sb_a\times\Sb_b)\equiv\pb_ {\As,1}$ (the form~(\ref{one})), is allowed in every one of the
symmetry operations.  It is robust, no symmetry can deny its existence as a contribution to the electric dipole
moment. The other part of $\pb_\As$, namely $\db_\perp\times(\Sb_a\times\Sb_b)\equiv\pb_{\As,2}$, plus the
contributions from the symmetric part, $\Ss$, of $\Cs$, have restrictions imposed by crystal symmetries that may
exist.

The other special contribution, $\pb\propto\Sb_a\times\Sb_b=\Wb$, discussed in the Introduction,  is seen to be
nonexistent if symmetries 3. or 4. exist. In general, contributions from $\sum \hat{\nu}C_{\nu\nu}W_\nu$
``contain" $\Wb$, but are not in its direction. Exceptions occur when $\Wb$ is in the x-direction (along AB),
and the symmetries present are 2., and/or 5., in which case $\pb \propto\Wb$ .

A few examples will illustrate the physical meaning of these single-bond results. \\
(a) Suppose the only symmetry is 2., Mirror $\perp$ AB, in which~(\ref{8}) holds. In this case we see that
$\db=\db_{||}$. Consider $\Wb$ in turn along the $x,y,z$ directions. $\Wb=\hat{x}: \pb=\hat{x}C_{xx};
\Wb=\hat{y}: \pb=\hat{y}C_{yy}+\hat{z}C_{zy}; \Wb=\hat{z}: \pb=\hat{y}C_{yz}+\hat{z}C_{zz}$. When $\Wb=\hat{y}$
or $\hat{z}$, the contribution from $\db\times\Wb$ is the z-component $C_{zy}$ or the y-component $C_{yz}$. That
there is no requirement that $\db=0$, i.e. $C_{yz}=C_{zy}$, makes sense, since symmetry 2. allows
the xy and xz planes to be nonequivalent.\\
(b) An example showing the new term $\db_\perp$: Suppose that the only symmetry is Mirror includes AB (3.).
Assume $\Wb=\hat{z}$. Then one can read off from~(\ref{9}) that $\pb=\hat{y}C_{yz}+\hat{x}C_{xz}.$ The
respective terms are $\propto\db_{||}\times\Wb$ and $\db_\perp\times\Wb$.\\
(c) An example relevant to the present literature is the dilemma posed in [50]: In the case of orbital
ordering considered by Jia et al~\cite{jia}, the bond symmetry is rather low; so why doesn't their calculation
yield one of the new forms, e.g. $\pb\propto \Sb_a\times\Sb_b$? The answer is given nicely by our results: The
d-orbitals at sites A and B are the $e_g$ states $3x^2-r^2$ and $3y^2-r^2$ respectively. Such a charge
configuration has the bond symmetries, reflection in plane containing AB (3.) and AB is a 2-fold axis (5.), and
only these. Looking at the corresponding $\Cs$ matrices (13) and the appropriately modified (14) for $n=2$, one
sees that the only possibility is $\pb=\db_{||}\times\Wb$. I.e. the \emph{particular} lowering of the bond
symmetry caused by orbital ordering is not sufficient to modify the form~(\ref{one}) for the dipole moment.
\vspace{.1in}\\
\textbf{Case 2}: $\textbf{i=j}$\\
Eq. (\ref{quad}) now becomes
\begin{equation}
\pb=\sum_{\gamma,\nu,\mu}\hat{\gamma}B_{\gamma
a\nu\mu}S_{a\nu}S_{a\mu}+\sum_{\gamma,\nu,\mu}\hat{\gamma}B_{\gamma b\nu\mu}S_{b\nu}S_{b\mu}.\label{i=j}
\end{equation}
Only the symmetric part, $B_{\gamma i\nu\mu}+B_{\gamma i\mu\nu}$ of $B_{\gamma i\nu\mu}$, for $i=a$ or $b$,
contributes. In order that this represent a canted-spin caused dipole, i.e. that it is zero for collinear spins
of arbitrary direction, one sees that
\begin{equation}
B_{\gamma a\nu\mu}+B_{\gamma b\nu\mu}=0.\nonumber
\end{equation}
That is, the part of~(\ref{i=j}) that gives a canted-spin-caused electric dipole is of the form
\begin{eqnarray}
\pb_0&=&\sum\hat{\gamma}B_{\gamma\nu\mu}(S_{a\nu}S_{a\mu}-S_{b\nu}S_{b\mu})\nonumber\\
   &\equiv&\sum\hat{\gamma}B_{\gamma\nu\mu}\Gamma_{\nu\mu},\label{i=jc}
\end{eqnarray}
where $B_{\gamma\nu\mu}=B_{\gamma a\nu\mu}$.
Clearly $\Gamma_{\nu\mu}=\Gamma_{\mu\nu}.$ It will be seen that this contains the form~(\ref{j}) as a special case.\\
We now apply the symmetry procedure to~(\ref{i=jc}).\\
\textbf{1. Inversion through O} \\
Again, $\Sb_a^\prime=\Sb_b,\Sb_b^\prime=\Sb_a$. Thus the right-hand side of~(\ref{i=jc}) changes sign, so
inversion invariance places no restriction on $B_{\gamma\nu\mu}$.\\
\textbf{2. Mirror $\perp$ AB} \\
From~(\ref{mirrorperp}) one readily sees that
\begin{equation}
\Gamma_{\nu\nu}^\prime=-\Gamma_{\nu\nu},\Gamma_{xy}^\prime=\Gamma_{xy},\Gamma_{xz}^\prime=\Gamma_{xz},
\Gamma_{yz}^\prime=-\Gamma_{yz}\nonumber
\end{equation}
Using these relations and demanding the vector property, $\pb_0^\prime=(-p_x,p_y,p_z)$ yields \\
\begin{widetext}
\begin{equation}
\left(\begin{array}{ccc}
B_{xxx} & 0 & 0\\
0 & B_{xyy} & B_{xyz}\\
0 & B_{xyz} & B_{xzz}
\end{array} \right),
\left(\begin{array}{ccc}
0 & B_{yxy} & B_{yxz}\\
B_{yxy} & 0 & 0\\
B_{yxz} & 0 & 0
\end{array} \right),
\left(\begin{array}{ccc}
0 & B_{zxy} & B_{zxz}\\
B_{zxy} & 0 & 0\\
B_{zxz} & 0 & 0          
\end{array} \right),
\end{equation}
\end{widetext}
where the 3 matrices represent $B_{\gamma\nu\mu}$ for $\gamma=x,y,z$, respectively, reading from left to
right.\\
\textbf{3. Mirror includes AB} \\
Taking the mirror as the xy plane, we have
\begin{equation}
\Gamma_{\nu\nu}^\prime=\Gamma_{\nu\nu}, \Gamma_{xy}^\prime=\Gamma_{xy} ,\Gamma_{xz}^\prime=-\Gamma_{xz},
\Gamma_{yz}^\prime=-\Gamma_{yz}.\nonumber
\end{equation}
This plus invoking the vector property of $\pb$ yields
\begin{widetext}
\begin{equation}
\left(\begin{array}{ccc}
B_{xxx} & B_{xxy} & 0\\
B_{xxy} & B_{xyy} & 0\\
0 & 0 & B_{xzz}
\end{array} \right),
\left(\begin{array}{ccc}
B_{yxx} & B_{yxy} & 0\\
B_{yxy} & B_{yyy} & 0\\
0 & 0 & B_{yzz}
\end{array} \right),
\left(\begin{array}{ccc}
0 & 0& B_{zxz}\\
0 & 0 & B_{zyz}\\
B_{zxz} & B_{zyz}&0   
\end{array} \right)
\end{equation}
\end{widetext}

\textbf{4. 2-fold rotation $\perp$ AB} \\
Taking the rotation axis as the z-axis this gives
\begin{equation}
\Gamma_{\nu\nu}^\prime=-\Gamma_{\nu\nu},\Gamma_{xy}^\prime=-\Gamma_{xy},\Gamma_{xz}^\prime=\Gamma_{xz},
\Gamma_{yz}^\prime=\Gamma_{yz},\nonumber
\end{equation}
which yields the identical form for the $B_{\gamma\nu\mu}$ matrices as (18).\vspace{.1in} \\
\textbf{5. n-fold axis along AB, n $\ge2$} \\
We again find that the form forced by rotation invariance depends on $n$. We discuss two examples, $n=2$ and 4.
In general, $\Gamma_{xx}^\prime=\Gamma_{xx}$ of course.

Beginning with $n=2$, we have
\begin{equation}
\Gamma_{\nu\nu}^\prime=\Gamma_{\nu\nu},\Gamma_{xy}^\prime=-\Gamma_{xy},\Gamma_{xz}^\prime=-\Gamma_{xz},
\Gamma_{yz}^\prime=\Gamma_{yz}.
\end{equation}
The form of the resulting $B_{\gamma\nu\mu}$ matrices is identical to (17).

For $n=4$, one readily finds that
\begin{equation}
\Gamma_{yy}^\prime=\Gamma_{zz},\Gamma_{zz}^\prime=\Gamma_{yy},\Gamma_{xy}^\prime=-\Gamma_{xz},
\Gamma_{xz}^\prime=\Gamma_{xy},\Gamma_{yz}^\prime=-\Gamma_{yz},
\end{equation}
which lead to
\begin{widetext}
\begin{equation}
\left(\begin{array}{ccc}
B_{xxx} & 0 & 0\\
0 & B_{xyy} & 0\\
0 & 0 & B_{xyy}
\end{array} \right),
\left(\begin{array}{ccc}
0 & B_{yxy} & B_{yxz}\\
B_{yxy} & 0 & 0\\
B_{yxz} & 0 & 0
\end{array} \right),
\left(\begin{array}{ccc}
0 & -B_{yxz}&B_{yxy} \\
-B_{yxz}& 0 & 0\\
B_{yxy} & 0&0
\end{array} \right)
\end{equation}
\end{widetext}
Comparison of the x-matrix with that in (17), which holds for $n$=2, shows that going from $n$=2 to the higher
symmetry $n=4$ gives the reduction $B_{xyy}-B_{xzz}\rightarrow 0$ and $B_{xyz}\rightarrow 0$. For the y and z
matrices the higher symmetry introduces no new zeros but brings in a relation between these matrices.

Finally, to compare with~(\ref{j}), we consider the case where all 5 symmetries hold, taking the case of 4-fold
rotation in symmetry 5.  We find the form of the B tensor is
\begin{widetext}
\begin{equation}
\left(\begin{array}{ccc}
B & 0 & 0\\
0 & C & 0\\
0 & 0 & C
\end{array} \right),
\left(\begin{array}{ccc}
0 & D & 0\\
D & 0 & 0\\
0 & 0 & 0
\end{array} \right),
\left(\begin{array}{ccc}
0 & 0&D \\
0& 0 & 0\\
D& 0&0
\end{array} \right).
\end{equation}
\end{widetext}
(Here $B=B_{xxx},C=B_{xyy},D=B_{yxy}$.) This gives
\begin{eqnarray}
\lefteqn{\pb=}\nonumber\\
& &\hat{x}(B-C)(S_{ax}^2-S_{bx}^2)+\nonumber\\
& &2D[\hat{y}(S_{ax}S_{ay}-S_{bx}S_{by})+\hat{z}(S_{ax}S_{az}-S_{bx}S_{bz})].\label{ourj}
\end{eqnarray}
 The corresponding term in~\cite{jia} (~(\ref{j}) in the present paper), is
\begin{eqnarray}
\lefteqn{\pb\propto\nonumber}\\
& =&\hat{x}(S_{ax}^2-S_{bx}^2)+ \nonumber\\
& &\hat{y}(S_{ax}S_{ay}-S_{bx}S_{by})+\hat{z}(S_{ax}S_{az}-S_{bx}S_{bz})
\end{eqnarray}
Thus it is seen that~(\ref{j})  is the special case of our result~(\ref{ourj}) where $B-C=2D$.

A particular case studied by~\cite{jia} applies to Mn$^{3+}$ as in the manganites, e.g. TbMnO$_3$, where the
t$_{2g}$ states are filled and the e$_g$ states are orbitally ordered (the spins on each ion are parallel). Jia
et al find no contribution of the form~(\ref{j}) whenever the t$_{2g}$ states with parallel spins are
filled~\cite{jia}. This fact motivates the application of our theory to this example. The e$_g$ orbitals on the
two sites are as described in example (c) under Case 1. The corresponding symmetry is: 2-fold axis along AB, and
two mirror planes, xy and xz . Applying our results to these cases we find
\begin{eqnarray}
\pb&=&\hat{x}\sum_{\nu=x,y,z} a_\nu(S_{a\nu}^2-S_{b\nu}^2)+\nonumber\\
& &\sum_{\nu=y,z} 2\hat{\nu}d_\nu(S_{ax}S_{a\nu}-S_{bx}S_{b\nu}),
\end{eqnarray}
where the $a_\nu$ and $d_\nu$ comprise 5 arbitrary coefficients.

Thus the symmetry does not require the vanishing of this type of contribution to $\pb$. This lack of generality
within the symmetry of the model~\cite{jia} indicates that other terms should enter. We suggest a candidate for such terms
is the modification of the spin-orbit coupling used in~\cite{jia} due to the presence of the O$^{2-}$ charge
near each Mn and the Mn$^{3+}$ charges near the oxygen ion. Such effects would not modify the symmetry of the
superexchange model of~\cite{jia}. (See also the related discussion in Section IV).

\textbf{III. Some applications to crystals (propagation of single-bond results).} Application of these local or
bond results requires their propagation to all other equivalent bonds. In this sense this approach becomes
``global", as is the powerful Landau theory of continuous phase transitions, also based in an essential way on
symmetry considerations. The approaches are, nevertheless, different. One aspect of the difference is that the
present theory applies to any phase of the crystal, whether or not it was reached through a continuous phase
transition from a known phase, unlike the Landau theory. Another symmetry approach, exemplified by the analyses
in~\cite{arima0,tokura}, considers the symmetry of the magnetically ordered crystal, and sees if that symmetry
is consistent with having a macroscopic electric polarization. In common with the present approach, its validity
is independent of how the phase was reached; it differs, e.g., in that it only considers the ferroelectric
response, whereas the present local symmetry approach allows prediction of
various complex anti-ferroelectric structures.  \\
\textbf{Case 1: $\mathbf{i\ne j}$}\\
The simplest application is a linear chain, spins in a line with no other objects around, as a check on
previously known results, given that the spins form a simple spiral. Here the $C$ matrix is the same for every
N.N. bond. In the usual case, the plane of the spins includes the chain direction, which is of course the
direction of the spiral wave vector. This sort of spiral, often called, appropriately, a cycloid, is actually
used to understand many real materials, e.g.~\cite{lawes,khomskii, cheong, katsura, arima0, arima, yamasaki2,
yamasaki, choi,malashevich}. But we can leave the direction of the spin plane (normal to $\Wb$) arbitrary for
the present discussion. In this case of high bond symmetry, every one of Moriya's symmetries applies.
Equations~(\ref{9}) and~(\ref{10}) imply
\begin{equation}
\Cs=\left(\begin{array}{ccc}
0 & 0 & 0\\
0 & 0 & C_{yz}\\
0 & -C_{yz} &  0 \end{array} \right).\label{11}
\end{equation}
Hence $\Cs$ is antisymmetric, $\db=\db_{||}$, so that $\pb=\db_{||}\times\Wb$. When $\Wb$ is in the z-direction
(spins lie in the x-y plane), this gives the expected result, $\pb$ in the y-direction. This is easily
generalized to 1-dimensional structures of lower symmetry by imagining the chain decorated with other charges;
in general each bond $\pb$ can, \'{a}  priori, be in any direction. If each decorated bond is just translated,
then the total $\pb$ will have other components. For example, if symmetries 3. and 4. are violated and 5.
remains, then $\Cs$ is given by~(\ref{10}) for $n>2$; in particular, if in addition $\Wb$ is in the x-direction,
then it follows that total $\pb$ is in the direction of $\Wb$. The same conclusion holds for $n=2$. This case is
that of Arima~\cite{arima0}, a ``proper screw" structure with $\pb$ in the direction of the spiral wave vector.
It is also related to the following.

The second example we discuss is  RbFe(MoO$_4$)$_2$ (RFMO), the ferroelectricity of which was studied
extensively by Kenzelmann et al~\cite{kenzelmann2}. While the observed ferroelectricity is well-understood by
the Landau-theory analysis of~\cite{kenzelmann2}, it is instructive to consider it from the point of view of the
present, quite different, symmetry theory. We consider the low-temperature behavior.

The magnetism resides in triangular layers of Fe$^{3+}$ ions whose spins lie in the planes, and form the
well-known $120^\circ$ spin order, which maintains the same handedness (the same $\Wb$ for each N.N. bond) in
translation from layer to layer.\footnote{One must remember that the arbitrary order taken in writing
$\Wb=\Sb_a\times\Sb_b$ makes sense only in conjunction with the associated $\stackrel{\Rightarrow}{\Cs}$. One
can make an assumption as to the order of the spins in $\Wb$, and thus the sign of $\pb$, for one bond. Then all
other bonds follow uniquely from crystal symmetry operations.} For the crystal structure see~\cite{kenzelmann},
particularly Fig. 1 a and b, and~\cite{inami}, particularly Fig. 1; the low-temperature (non-magnetic) space
group is P$\bar{3}$. Other non-magnetic ions between these layers cause the symmetries 3. and 4. to be violated.
Whether or not any of the remaining symmetries exist, it is seen that a local electric dipole moment $\propto
\Wb$, which lies $\perp$ these planes, is allowed. Each plane $\nu$ possesses a total dipole moment $\Pb_\nu$,
as follows from the 3-fold axis of P$\bar{3}$ which implies that $\Cs$ for every bond within a plane is rotated
by this operation. Also, the 120$^\circ$ spin structure has the same property. Further, we need to know if all
planes produce the same moment, or might the sign alternate. Now P$\bar{3}$ implies a center of inversion
between the magnetic planes that connect bonds in different planes, carrying all the complex non-magnetic
structure along via the inversion. Essential is the relation between the $\Cs$ matrices describing the
surroundings of each of the inversion-related bonds. We determine this as follows. We have
$\pb=\stackrel{\Rightarrow}{\Cs}\cdot\Wb$, so that $I\pb=I\stackrel{\Rightarrow}{\Cs}I\cdot I\Wb.$ But
$I\Wb=-\Wb$, as noted above. Since $I\pb=-\pb$, it follows quite generally, that
\begin{equation}
I\stackrel{\Rightarrow}{\Cs}I=\stackrel{\Rightarrow}{\Cs};\label{12}
\end{equation}
i.e., $\Cs$ is invariant under inversion. $\Wb$ being the same for every plane, it follows that the planar
$\Pb_\nu$'s all have the same sign, resulting in a net non-zero polarization, as observed.
\begin{figure}[hh]
\includegraphics[height=2.in]{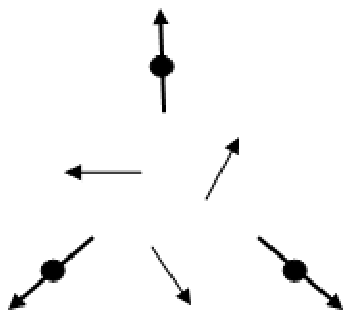}\vspace{-60pt}
\centering
 \caption{Triangular plaquette of spins (darker arrows) and electric dipoles (lighter arrows) predicted for RFMO (schematic).
 The arrows represent the projections on the spin planes of the full dipoles.}
\label{fig5}
\end{figure}

The authors note~\cite{kenzelmann2} that the existence of a 3-fold axis $\perp$ to the planes (the c-axis)
implies there can not be a component of $\Pb$ parallel to the planes. We can see this from our general
expression, $\pb=\hat{x}\Cs_{xz}+\hat{y}\Cs_{yz}+\hat{z}\Cs_{zz}$, for our case, $\Wb\propto\hat{z}$: For each
triangular plaquette, the x- and y- components will add to zero because of the 3-fold axis. On the other hand,
these components $\perp \hat{z}$ will order \emph{antiferroelectrically} in a $120^\circ$ state because of the
ordered spins and the 3-fold axis. If the high-T structure, space group P$\bar{3}m1$, held, then only the
$\db_{||}$ term would survive, and that would imply that the projection of the bond dipole moments would each
lie $\perp$ to the bond. Fig. 5 in an early effort~\cite{kaplanarxiv} shows this for a triangular plaquette.
However, the true structure has the lower symmetry space group $P{\bar{3}}$; one can see (particularly with the
help of Fig. 1 in~\cite{inami}) that none of Moriya's symmetry operations holds, so that any direction of $\pb$
for given spins in a bond is allowed by symmetry.  We indicate this situation schematically for a single
triangular plaquette in Fig. 2. The location of the electric moments at the midpoints of the triangle edges (the
Kagom\'e structure, dual to the triangular lattice) is symbolic of the actual bond charge density found in the
microscopic theories of~\cite{katsura,jia,kaplan} (although, with the exception of~\cite{kaplan}b, the high
symmetry assumed in these calculations requires no component of $\pb ||\Wb$). Such a charge distribution would
be ordered in the crystal (it's tied strongly to the magnetism), and would induce corresponding changes in ionic
positions, which should help in its detection by diffraction methods.

Also the response of the multiferroic state to a uniform magnetic field $\Hb$ might possibly give insight into
this complex orientation structure of the local dipoles. The idea is, of course, that applying $\Hb$ will
distort the magnetic order, modifying $\Wb$ and therefore the local dipoles $\pb_\nu$. This idea was discussed
in~\cite{delaney}. In particular, applying $\Hb$ in the plane of the spins in FIG. 2 would give a net dipole
moment for the plaquette, considering the system of 3 spins as isolated.  We have shown that for  small $\Hb\
||$ to one of the spins, the component of total polarization in an isolated triangular lattice with the
120$^\circ$ spin structure is of order $H^2$. In RFMO there have appeared some limited experimental studies of
the magnetic and electric (i.e. charge) properties in applied fields~\cite{kenzelmann2}. These put $\Hb$
parallel to the plane of the spins along a particular crystallographic direction, and presented information
about the c-axis component of polarization, only as to whether it was zero or non-zero. The theory presented was
for the zero-field case. In fact the theory for $\Hb\ne0$ is non-existent as far as we are aware, and that is
essentially because the particular magnetic structure in a field is complex and its origin has not been
elucidated, particularly concerning the incommensurate component of the spiral wave vector along the
c-axis~\cite{svistov}. See also~\cite{chubukov}. While such studies would be interesting, we won't consider them
here.

Our last example concerns the materials CuFeO$_2$ and ACrO$_2$ (A=Cu,Ag), in which canted-spin-caused
ferroelectricity was found~\cite{kimura2,nakajima,seki0,seki,soda}. These materials have the magnetic ions
(Fe$^{3+},$ Cr$^{3+}$) situated on triangular lattices (basal planes), and are of delafossite form. The canted
spin states are spirals with wave vector $\Qb=(q,q)$ in the plane and the spins lie in a plane such that
$\Sb_a\times\Sb_b$ lies in the basal plane. The special case $\Sb_a\times\Sb_b||\Qb$ is known to occur in
CuFeO$_2$ and CuCrO$_2$~\cite{nakajima, soda}. Importantly, in the latter cases the polarization lies parallel
to $\Qb$, i.e. in the direction $\Sb_a\times\Sb_b$, where the two spins are N.N.'s along the $\Qb$ direction. As
discussed above, there is a close relation between this structure and that of RFMO: the essential difference is
that the magnetic anisotropy is easy plane for RFMO or easy axis for the former, as emphasized
in~\cite{seki,tokura}. But in all these cases, the polarization is in the direction of $\Sb_a\times\Sb_b$. We
just saw how our symmetry analysis gives results consistent with these facts for RFMO. Let's consider now the
delafossites. Referring to Fig. 1 of Arima's paper~\cite{arima0}, one sees that the only one of the 5 symmetry
operations that is satisfied for a N.N. Fe-Fe bond is a 2-fold rotation axis coinciding with AB (operation 5.),
for which the C-matrix is given by~(\ref{10}) appropriately modified for $n=2$, where the bond is along the
x-direction. But $\Sb_a\times\Sb_b=\Wb$ is also in the x-direction, giving $\pb=C_{xx}\Wb$ (in the x-direction),
i.e. $\pb$ is in the direction of $\Qb$ as observed.

\textbf{IV Concluding remarks}

The robustness of $\pb_{\As,1}=\Rb\times(\Sb_a\times\Sb_b)$  under symmetry requirements may be why it has been
found experimentally in many different materials, whereas only one of the many other possibilities given by the
present theory has been found, as far as we are aware, namely $\pb$ in the direction of $\Sb_a\times\Sb_b$, and
only in three materials, namely CuFeO$_2$~\cite{nakajima}, CuCrO$_2$~\cite{soda} and RbFe(MoO$_4$)$_2$
(RFMO)~\cite{kenzelmann2}.

$\pb_{\As,2}$ shows new possibilities for the dipole moment produced by a pair of atoms with canted spins. E.g.
in the case familiar in many multiferroics where $\Sb_a,\Sb_b,$ and $\Rb$ are coplanar, say in the x-y plane,
then the already discovered possibility, $\pb$ has a y-component (from $\pb_{\As,1}$), is now accompanied by the
possibility of having a z-component originating from $\pb_{\As,2}$. There can also be an x-component ($||
\Sb_a\times\Sb_b$), originating from $\pb_\Ss$.

The results obtained here apply directly to model calculations based on clusters that contain a pair of magnetic
atoms, as in~\cite{katsura,jia,sergienko}. The process of checking our symmetry results against the simple,
idealized quantum-mechanical model~\cite{kaplan} described in Appendix 2 goes further in that it suggests a
microscopic mechanism for the case where the dipole moment $\pb$ is in the direction of $\Sb_a\times\Sb_b$,
which includes both the proper screw structure~\cite{arima0} and the spiral in RFMO~\cite{kenzelmann}. The
mechanism, that should be valid in the approach of~\cite{katsura,jia,sergienko},  is the effect of the
environment on the nature, or symmetry, of the spin-orbit interaction. The SO interaction in an isolated atom or
ion is of the commonly used form $\propto \mathbf{l}\cdot\mathbf{s}$, and this is the form used in the theories
of~\cite{katsura,jia}. However this is just the special case of the more general form $\propto \nabla
V(\rb)\times \tilde{\pb}\cdot\mathbf{s}$ ($\tilde{\pb}$ is the momentum operator) that results when $V(\rb)$ is
spherically symmetric, as assumed for the nucleus plus the other electrons on the atom. When the atom is in an
environment of other charges outside the atom, $V(\rb)$ will have a non-spherically-symmetric
part.~\cite{pederson} This will reflect the symmetry of that environment and will lead to the other forms of the
magnetically induced electric dipole.~\footnote{This effect is implicit in the analysis of~\cite{sergienko} via
the microscopic theory behind the DM vector $\Db$~\cite{moriya}.} This mechanism differs substantially from
Arima's~\cite{arima0}: this is linear in the SO coupling strength, whereas Arima's is 2nd order. Of course, this
contribution will generally be smaller than the intra-atomic (spherical) contribution, because the environmental
charges are farther from the atom than the atomic charge, an effect ameliorated by the fact that the active
electron states vanish at the nucleus, both for the magnetic ions and the oxygen. A crude estimate suggests that
this mechanism is not negligible compared to the spherical term, originating in the d-shell.

Our results of course suggest strongly that there will be materials that exhibit the new forms for $\pb$. We
have given three examples of the single form $\mathbf{p}\propto\mathbf{S}_a\times\Sb_b$, namely CuFeO$_2$,
CuCrO$_2$ and RFMO. The observation of others would be of great interest in verifying the theory and deepening
our understanding of these fascinating multiferroics.

We note that the present local or bond-symmetry approach can also be applied to the symmetric magnetostriction
(the tensor $B_{\gamma\nu\mu}^s$ defined in Appendix 1), which would include electric dipoles produced by
collinear magnetic ordering.

We thank Mr. Z. Rak and Dr. Mal-Soon for help in understanding the RFMO structure. Helpful communications with
A. B. Harris, G. Lawes, M. Kenzelmann, Y. Tokura, N. Nagaosa, H. Katsura, A. V. Balatzky, N. Furukawa, and C.
Jia are gratefully acknowledged.

\textbf{Appendix 1: Proof that Equation~(\ref{3}) is the most general vector function of spins $\Sb_a,\Sb_b$,
bilinear in
the spins, and representing ``canted-spin-caused" electric dipoles}\\

The most general vector function of spins bilinear in the spins, $\Sb_a,\Sb_b$ is
\begin{equation}
\mathbf{p}=\sum_{\gamma,\nu,\mu}\hat{\gamma}B_{\gamma\nu\mu}S_{a\nu}S_{b\mu},
\end{equation}
where $\gamma,\nu,\mu$ run over Cartesian components $x,y,z$. The spins are assumed to be of fixed length, so
they can be taken as unit vectors (or, really, unit pseudovectors, but this is irrelevant here). The idea that
$\mathbf{p}$ be ``caused" by spin canting is defined by the requirement $\mathbf{p}=0$ if $\Sb_a=\pm\Sb_b$ for
arbitrary $\Sb_a$. I.e., $\mathbf{p}$ vanishes whenever the spins are collinear (non-canted).

We can write
\begin{equation}
B_{\gamma\nu\mu}=B_{\gamma\nu\mu}^s+B_{\gamma\nu\mu}^a,
\end{equation}
where $B_{\gamma\nu\mu}^s=B_{\gamma\mu\nu}^s, B_{\gamma\nu\mu}^a= -B_{\gamma\mu\nu}^a$, defining in the obvious
correspondence $\mathbf{p}_s$ and $\mathbf{p}_a$, with $\mathbf{p}=\mathbf{p}_s+\mathbf{p}_a$. See footnote
[53]. It can be verified straightforwardly that
\begin{equation}
\mathbf{p_a}=\sum_{\gamma\beta}\hat{\gamma}C_{\gamma\beta}(\Sb_a\times\Sb_b)_\beta,\label{15}
\end{equation}
where
\begin{eqnarray}
C_{\gamma x}&=&B_{\gamma yz}^a\nonumber\\
C_{\gamma y}&=&B_{\gamma zx}^a\nonumber\\
C_{\gamma z}&=&B_{\gamma xy}^a.
\end{eqnarray}
Clearly $\mathbf{p}_a=0$ for collinear spins, and~(\ref{15}) for $\mathbf{p}_a$ is the same form as~(\ref{3})
for $\pb$.

Now consider the symmetric component. Putting $\Sb_a=\pm\Sb_b$, we have
\begin{equation}
\mathbf{p}_s=\pm\sum \hat{\gamma}B^s_{\gamma\nu\mu}S_{a\nu}S_{a\mu}
\end{equation}
Choosing $\Sb_a$, in turn, along the x,y,z directions, and in the xy,yz,zx planes one sees that
\begin{equation}
\mathbf{p}_s=0 \ \mbox{for all}\  \Sb_a \ \mbox{implies}\  B^s_{\gamma\nu\mu}=0\ \mbox{for all}\ \gamma,\mu,\nu.
\end{equation}
This then proves that~(\ref{3}) uniquely embodies the idea of canted-spin-caused electric dipoles (within the
assumption of a bilinear form). It also implies that for any moment resulting from the symmetric component, any
canting is incidental, i.e. non-essential.

\textbf{Appendix 2: Simple microscopic model for canted-spin-caused electric dipole}

The basic model for the calculations in~\cite{kaplan}, generalized to arbitrary symmetry of the bond plus its
surroundings (``the crystal"), is presented here. Illustration of its use for checking the abstract symmetry and
propagation operations is given.

We consider two essentially one-electron atoms, e.g., 2 hydrogens, or 2 lithiums. The generalization to two
different alkali atoms is not difficult, but for simplicity is not given here. There are 8 spatial wave
functions in the basis, an s and 3 p-states for each atom. The average spins on each site (A and B, as in Fig.
1) are fixed so that the 1-electron basis has just 8 states. We write these as
\[s_a\chi_a,s_b\chi_b,p_{a\nu}\chi_a,p_{b\nu}\chi_b, \nu=x,y,z,\]
 where $\chi_a$ and $\chi_b$ are the spin
states. The spatial parts are assumed to be Wannier functions, i.e. they are hybridized to make them mutually
orthogonal (the overlaps of atomic orbitals are assumed small).  We denote the 2 s-states as $\phi_i, i=1,2$,
the remaining states as $\phi_i,i=3,\cdots 8$. So each unperturbed atom has two energies, the s-state and the
p-state, separated by $\Delta_0>0$. The model Hamiltonian is
\begin{equation}
H=\Delta_0\sum_{i=3}^8 n_i+(\tilde{\sum}_{i\le2,j>2} v_{ij}c_i^\dagger
c_j+\mbox{h.c.})+U\sum_{\mbox{on-site}}n_i n_j. ~\label{13}
\end{equation}
Here $n_i=c_i^\dagger c_i$ and $\tilde{\sum}$ means to sum only over terms where $i$ and $j$ refer to different
sites. Sample terms are $<s_a\chi_a|v|p_{b\nu}\chi_b>$. $v$ is the spin-orbit coupling operator
\begin{equation}
v=a_0\nabla V\times \tilde{\pb}\cdot\mathbf{s},\label{14}
\end{equation}
where $V$ (which appears in~\cite{moriya}) is an effective potential energy that reflects the crystal symmetry
excluding magnetic ordering and spin-orbit coupling, $a_0$ involves only fundamental constants, and
$(\tilde{\pb},\mathbf{s})=(\mbox{momentum}/\hbar,\mbox{spin}/\hbar)$.

In fact, the 1-electron operator~(\ref{14}) is an approximation to the actual spin-orbit coupling which is a
rather complicated 2-electron operator.~\cite{bethe} There is a considerable literature attempting to calculate
SO effects in various approximation schemes, e.g. Hartree-Fock approximation~\cite{blume1,blume2,blume3}
considering single atoms, a different mean field approximation~\cite{neese} applicable to many-center systems.
The latter found that a local potential gave excellent results for g-tensors in certain molecules (although in
the best approximation $V$ is non-local). The simplest approximation that we found in the literature used the
Coulomb or Hartree term for $V(\rb)$~\cite{pederson}. We explicitly make use of locality, and the only property
important for the present considerations is that it be true to the symmetry of the system studied.

Because the spin-orbit term includes only transitions, $s_a\rightarrow p_{b\nu}$ and $s_b\rightarrow p_{a\nu}$
between the two sites, we call this inter-site spin-orbit coupling.

The unperturbed ground state for the system is
\begin{equation}
\Phi_0=c_1^\dagger c_2^\dagger|0>.
\end{equation}
To first order, the perturbed ground state is
\begin{equation}
\Phi=\Phi_0-\Delta^{-1}\tilde{\sum}_{i\le2,j>2} v_{ji}c_j^\dagger c_i\Phi_0,
\end{equation}
where $\Delta=\Delta_0+U$. Measuring $\rb$ from the mid-point of the bond, it easily shown that
$<\Phi_0|\rb_1+\rb_2|\Phi_0>=0$, so that the electric dipole moment
\begin{eqnarray}
\pb&=& e<\Phi|\sum_{ij}<\phi_i|\rb|\phi_j>c_i^\dagger c_j|\Phi>\nonumber\\
&=&-\frac{e}{\Delta}\tilde{\sum}_{i\le2,j>2} v_{ji}<i|\rb|j>+\mbox{c.c.}
\end{eqnarray}
to leading order. In terms of the explicit 1-electron states this is
\begin{eqnarray}
\pb&=&-\frac{e}{\Delta}\sum_\nu<p_{a\nu}\chi_a|v|s_b\chi_b><s_b\chi_b|\rb|p_{a\nu}\chi_a>\nonumber\\
& & +(a\leftrightarrow b)+\mbox{c.c.}.\label{18}
\end{eqnarray}
We have
\begin{eqnarray}
<p_{a\nu}\chi_a|v|s_b\chi_b>&=&-i a_0<p_{a\nu}|\nabla V\times\nabla|s_b>\nonumber\\
& &\cdot<\chi_a|\mathbf{s}|\chi_b>\nonumber\\
<s_b\chi_b|\rb|p_{a\nu}\chi_a>&=&<s_b|\rb|p_{a\nu}><\chi_b |\chi_a>.\label{19}
\end{eqnarray}
With
\begin{eqnarray}
\mathbf{u}&\equiv&<\chi_a|\mathbf{s}|\chi_b>\nonumber\\
w&\equiv&<\chi_a|\chi_b>,\label{20}
\end{eqnarray}
and~(\ref{19}), (\ref{18}) becomes
\begin{eqnarray}
\pb&=&-2a_0\frac{e}{\Delta}\sum_\nu<p_{a\nu}|\nabla V\times\nabla
s_b>\cdot\mbox{Im}(\mathbf{u}w^*)<s_b|\rb|p_{a\nu}>\nonumber\\
& & +(a\leftrightarrow b)\label{21}
\end{eqnarray}
With the help of the well-known equations for $\chi_c$ such that the average in $\chi_c,
<\mathbf{s}>_c\equiv\Sb_c$ points in the direction with polar angles $\theta_c,\phi_c$, one can show that
\begin{equation}
\mbox{Im}(\mathbf{u}w^*)=-\Sb_a\times\Sb_b.~\label{22}
\end{equation}
We then obtain
\begin{eqnarray}
\pb&=&2a_0\frac{e}{\Delta}\sum_\nu[<s_b|\rb|p_{a\nu}><p_{a\nu}|\nabla V\times\nabla
s_b>\nonumber\\
& &-(a\leftrightarrow b)]\cdot\Sb_a\times\Sb_b.~\label{23}
\end{eqnarray}

Choose the p-functions as
\begin{eqnarray}
p_{a\nu}&=&\nu p_a, \nu=y,z\nonumber\\
p_{b\nu}&=&\nu p_b, \nu=y,z\nonumber\\
p_{a x}&=&(x+1/2)p_a\nonumber\\
p_{b x}&=&(x-1/2)p_b,\label{24}
\end{eqnarray}
where $p_a,p_b$ are spherically symmetric about points A,B respectively.
 Define
\begin{equation}
\mathbf{T}_{ba}^\nu\equiv<s_b|\rb|p_{a\nu}>=\hat{\nu}T_{ba}^\nu.\label{25}
\end{equation}
The last equality follows from the cylindrical symmetry of $s_b(\rb)p_a(\rb)$. Further
\begin{eqnarray}
T_{ba}^\nu&=&<s_b|\nu^2p_a>=T_{ab}^\nu\ \ \mbox{for}\ \nu=y,z\nonumber\\
T_{ba}^x&=&<s_b|x(x+1/2)p_a>=T_{ab}^x,\label{26}
\end{eqnarray}
These results follow from $p_{a\nu}(-x,y,z)=p_{b\nu}(x,y,z)$ for $\nu=y,z$ and $p_{a x}(-x,y,z)=-p_{b
x}(x,y,z)$. Hence the quantity $<s_b|\rb|p_{a\nu}>$ factors out of the square bracket in~(\ref{23}).

Comparison of~(\ref{23}) with~(\ref{3}) shows that, to within the constant factor $2a_0e/\Delta$, the basic
matrix defined in the general theory~(\ref{3}) is
\begin{eqnarray}
C_{\nu\mu}&=&T_{ba}^\nu[<p_{a\nu}|(\nabla V\times\nabla)_\mu s_b>-(a \leftrightarrow b)]\nonumber\\
&\equiv&T_{ba}^\nu(I_{ab}^{\nu\mu}-I_{ba}^{\nu\mu})\label{27}
\end{eqnarray}
for the present detailed microscopic model.

Let us first check the fundamental result~(\ref{12}) that $C_{\nu\mu}$ is invariant under inversion. We
calculate $C^\prime$ the inverted $C_{\nu\mu}$ by replacing $V(\rb)$ by $V^\prime=V(-\rb)$. Consider e.g.
\begin{eqnarray}
C_{xx}^\prime&=&T_{ba}^x[\int d^3r\ p_{a x}\nonumber\\
& &(\frac{\partial V^\prime}{\partial y}\frac{\partial s_b}{\partial z}-\frac{\partial V^\prime}{\partial
z}\frac{\partial s_b}{\partial y})-(a\leftrightarrow b)].
\end{eqnarray}
On changing the integration variables $\rb\rightarrow-\rb$, $V^\prime\rightarrow V$ and $I_{ab}^{xx}\rightarrow
-I_{ba}^{xx}$, so that $C_{xx}^\prime=T_{ba}^x(-I_{ba}^{xx}+I_{ab}^{xx})$, which is $C_{xx}$. The property $p_{a
x}(-x,y,z)=-p_{b x}(x,y,z)$ was essential to the conclusion.

Now consider checking some of the symmetry rules corresponding to Moriya's 5 symmetry operations. \\
\textbf{Rule 1.} $C_{\nu\mu}$ doesn't change under inversion whether or not the system is invariant under
inversion, as
was just shown. Hence the general conclusion, inversion symmetry places no restriction on $C$, is verified for the model.\\
\textbf{Rule 2.} Here $V(x,y,z)=V(-x,y,z).$ Thus, e.g., in $C_{xx}$ the integral $I_{ab}^{xx} =-I_{ba}^{xx}$,
seen by changing the integration variable $x$ to $-x$, returning the initial expression. I.e., this symmetry
puts no restriction on $C_{xx}$. Next,
\begin{eqnarray}
C_{xy}&=&T_{ba}^x[\int d^3r p_{ax}(\frac{\partial V}{\partial z}\frac{\partial s_b}{\partial
x}-\frac{\partial V}{\partial x}\frac{\partial s_b}{\partial z})-(a\leftrightarrow b)]\nonumber\\
&=&T_{ba}^x(I_{ba}^{xy}-I_{ab}^{xy})=-C_{xy}.
\end{eqnarray}
(One sees that $I_{ab}^{xy}=I_{ba}^{xy}.$) Therefore $C_{xy}=0$. Thus the model has verified two of the matrix
elements in~(\ref{8}), deduced earlier by a general, model-independent, symmetry argument. These examples should
suffice to illustrate the procedure, which can be seen to check all the previous results.

\end{document}